\documentclass[9pt,twocolumn,twoside]{osajnl}
\usepackage{xcolor}
\usepackage[normalem]{ulem}

\journal{ol} 

\setboolean{shortarticle}{true}

\begin{document}

\title{Biphoton engineering using modal spatial overlap on-chip}

\author[1]{Xiangyan Ding}
\author[1]{Jing Ma}
\author[1]{Liying Tan}
\author[2]{Amr S. Helmy}
\author[1,*]{Dongpeng Kang}

\affil[1]{School of Astronautics and National Key Laboratory of Science and Technology on Tunable Laser, Harbin Institute of Technology, 92 West Dazhi Street, Harbin, 150001, China}
\affil[2]{The Edward S. Rogers Department of Electrical and Computer Engineering, University of Toronto, 10 King’s College Road, Toronto, Ontario, Canada M5S 3G4}

\affil[*]{Corresponding author: dongpeng.kang@hit.edu.cn}

\dates{Compiled \today}


\doi{\url{http://dx.doi.org/10.1364/XX.XX.XXXXXX}}

\begin{abstract}
Photon pairs generated by spontaneous parametric down-conversion are essential for optical quantum information processing, in which the quality of biphoton states is crucial for the performance. To engineer the biphoton wavefunction (BWF) on-chip, the pump envelope function and the phase matching function are commonly adjusted, while the modal field overlap has been considered as a constant in the frequency range of interest. In this work, by utilizing modal coupling in a system of coupled waveguides, we explore the modal field overlap as a new degree of freedom for biphoton engineering. We provide design examples for on-chip generations of polarization entangled photons and heralded single photons, respectively. This strategy can be applied to waveguides of different materials and structures, offering new possibilities for photonic quantum state engineering.
\end{abstract}

\setboolean{displaycopyright}{true}

\maketitle

Quantum states of photons, including entangled photons and single photons, are essential sources for optical quantum information processing, such as communications, computation, simulation, and metrology \cite{o2009photonic}. In particular, spontaneous parametric down-conversion (SPDC) is a convenient process to generate photon pairs for sources of photonic quantum states \cite{doi:10.1063/5.0030258}. In SPDC, a pump (P) beam illuminates a nonlinear medium with $\chi^{\left(2\right)}$ nonlinearity, and through the interaction obeying energy conservation, there is a probability that a pump photon will split into two new photons, namely signal (S) and idler (I), being entangled or uncorrelated. The corresponding biphoton wavefunction (BWF) reads $\psi(\omega_S,\omega_I)\propto\alpha\left(\omega_S+\omega_I\right)\phi\left(\omega_S,\omega_I\right)$, where $\alpha\left(\omega_S+\omega_I\right)$ is the pump envelope function (PEF) and $\phi\left(\omega_S,\omega_I\right)$ is the phase-matching function (PMF), usually in the form of a sinc function.

However, photon pairs generated by SPDC usually do not directly meet various requirements for different applications. These requirements include the suppression of distinguishing information \cite{PhysRevLett.83.955}, the control of photon correlations \cite{Wei:21}, the need for narrowband entangled states for minimizing the chromatic dispersions \cite{RevModPhys.74.145}, and the need for broadband states for resolution enhancement in quantum sensing \cite{Chen:21}, etc. To achieve a high performance in optical quantum information applications, the BWF of photon pairs often needs further shaping before they can be used. Some extra components have been applied to shape the biphoton-state, such as time lenses \cite{PhysRevLett.117.243602}, dispersive elements \cite{Jin:18}, or filters \cite{Kysela26118}, but they inevitably increase the complexity of the photonic circuits. As such, shaping the biphoton-state is preferable to take place directly at the generation stage.

According to the expression of BWF mentioned above, its shaping relies solely on either adjusting the PEF or the PMF. In the former case, the PEF is uniquely determined by the pump laser. However, in the latter case, while the PMF is primarily determined by the material property, a variety of techniques have been developed to adjust its shape. These include dispersion engineering that either adjusts group velocities \cite{Saravi:19,Svozilik:11} or group velocity dispersions \cite{PhysRevLett.127.183601} of the interacting photons, domain-engineering for tailoring the nonlinearity of crystals \cite{Pickston:21}, and pump spatial engineering \cite{Francesconi:20}, etc. However, generic and versatile techniques for shaping the BWF are still of particular interest.

In this work, we exploit the possibility of using the modal field overlap as a new degree of freedom to shape the BWF. While the spatial overlap of the interacting fields is a function of wavelength in theory, it is essentially a constant over the whole frequency range of interest in all previous work, to the best of our knowledge. Here we utilize modal coupling in asymmetric heterogeneously coupled waveguides so that the spatial overlap could vary significantly within the bandwidth of a BWF, acting as an intrinsic spectral filter. This technique can be applied to waveguides of different materials and used in conjunction with previously developed techniques to achieve more flexible biphoton engineering. For illustrations, design examples based on two completely different, yet readily available platforms, i. e., \uppercase\expandafter{\romannumeral3}–\uppercase\expandafter{\romannumeral5} semiconductor and Lithium Niobate on insulator (LNOI) waveguides, are provided to show the improvements of polarization entanglement and spectral purity, respectively.

\begin{figure}[bt!]
	\centering\includegraphics[width=\columnwidth]{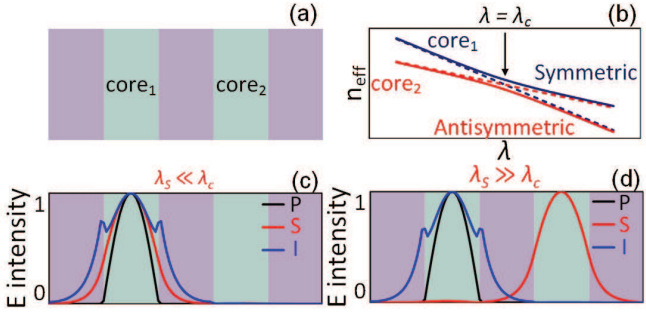}
	\caption{(a) Schematic illustration of the five-layer slab waveguide, and (b) its $n_{\text{eff}}$ of symmetric and antisymmetric supermodes. Dashed lines are the $n_{\text{eff}}$ of two isolated waveguide modes. (c) The electric intensity profiles of the pump, signal, and idler modes when $\lambda_S\ll\lambda_c$, and (d) when $\lambda_S\gg\lambda_c$.}
	\label{fig1}
\end{figure}
For simplicity, we first consider the coupling between two slab modes in a one-dimensional five-layer slab waveguide, as shown in Fig. \ref{fig1} (a). We assume that the two core layers (core$_1$ and core$_2$) are made of different materials and their refractive indices are both larger than that of the surrounding layers. The effective indices ($n_{\text{eff}}$) of the two isolated modes corresponding to core$_1$ and core$_2$ intersects at the coupling wavelength of $\lambda_c$. The two modes couple with each other and form a pair of symmetric and antisymmetric supermodes. The $n_{\text{eff}}$ of the supermodes asymptotically approach those of the isolated waveguide modes when the wavelength moves away from $\lambda_c$, as shown in Fig. \ref{fig1} (b)(See supplementary 1 for more details).

The $n_{\text{eff}}$ is an indicator for the modal field profile, which also shows the same asymptotic behaviors. To illustrate this phenomenon, the electric intensity distributions at two different signal-idler wavelength pairs with the same pump are shown in Fig. \ref{fig1} (c) and Fig. \ref{fig1} (d) respectively. Here we assumed the signal photon is generated in the TE polarized symmetric supermode while idler in the uncoupled TM mode. Note that photons are generated in waveguide spatial modes rather than any particular layer. For the case shown in Fig. \ref{fig1} (c), the signal wavelength is sufficiently lower than the coupling wavelength, i. e., $\lambda_S\ll\lambda_c$, and all three interacting fields locate predominately in layer core$_1$. While for the latter case shown in Fig. \ref{fig1} (d), in which $\lambda_S\gg\lambda_c$, the location of the idler field keeps unchanged but that of the signal field moves to layer core$_2$, indicating a next to zero overlap with three interacting modes. These cases show the best and worst scenarios of modal overlaps, and any wavelength in between is of no particular importance in our analysis.

The spectral range in which the modal profile changes depends on how strong two waveguides couple. For strongly coupled waveguides, the modal profile gradually changes in a broad spectral range, and thus the three-field overlap can be considered constant \cite{Ding:21}. However, for weak coupling, this overlap varies rapidly within the wavelength range of interest. It affects the pair generation rate at given wavelengths and can be quantified by the effective area $A(\omega_S,\omega_I)$, given by \cite{PhysRevA.77.033808}
\begin{equation}
\begin{split}
\frac{1}{\sqrt{A(\omega_S,\omega_I)}}\propto&\int dxdy \chi_{(2)}^{\alpha\beta\gamma}(x,y)e_S^\alpha(x,y)e_I^\beta(x,y)\left[ e_P^\gamma(x,y) \right]^*,
\label{EqA}
\end{split}
\end{equation}
where $\mathbf{e}_a(x,y)$ ($a=(S,I,P)$) is the modal electric field profile, and $\chi_{(2)}^{\alpha\beta\gamma}(x,y)$ is the structural nonlinearity distribution. Here $\alpha$, $\beta$ and $\gamma$ stand for Cartesian coordinates and summations over repeated indices are omitted. In this case, the BWF needs to be modified to
\begin{equation}
\psi(\omega_S,\omega_I)\propto\sqrt\frac{1}{A(\omega_S,\omega_I)}\alpha\left(\omega_S+\omega_I\right)\phi\left(\omega_S,\omega_I\right).
\label{Eq4}
\end{equation}
As such, $A(\omega_S,\omega_I)$ is equivalent to a spectral filter and can be used as a new degree of freedom to engineer the BWF.

\begin{figure}[bt!]
	\centering\includegraphics[width=\columnwidth]{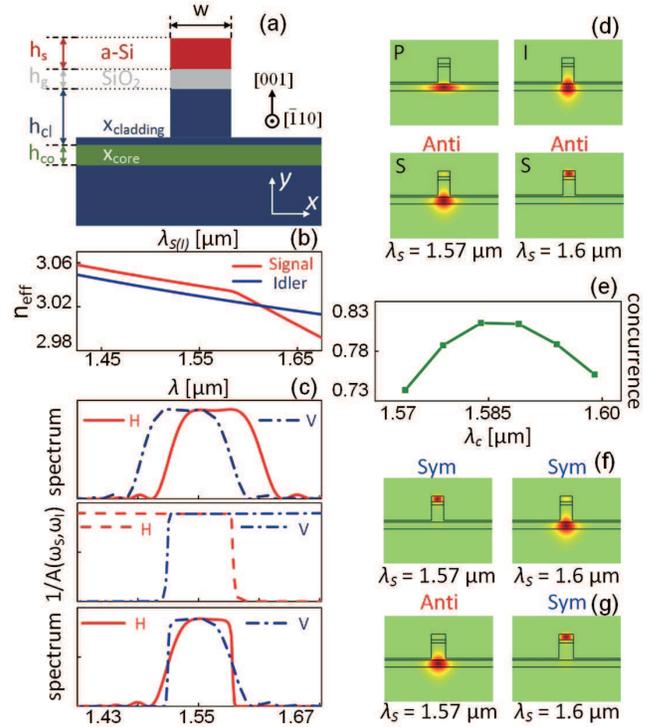}
	\caption{(a) Schematic cross-sectional view of an Al$_x$Ga$_{1-x}$As-a-Si coupled ridge waveguide. (b) The $n_{\text{eff}}$ of the TE antisymmetric supermode and uncoupled TM mode. (c) Spectra of down-converted photons in the isolated Al$_x$Ga$_{1-x}$As waveguide (upper row), wavelength dependences of $1/A(\omega_S,\omega_I)$ if Al$_x$Ga$_{1-x}$As is coupled with a-Si (middle row), and spectra of down-converted photons in the coupled waveguide (lower row). (d) Modal field profiles of the pump, idler, and antisymmetric signal at $\lambda_S<\lambda_c$ and $\lambda_S>\lambda_c$. (e) The dependence of concurrence on the coupling wavelength $\lambda_c$. (f) The unwanted symmetric supermode profiles at different wavelengths. (g) The antisymmetric and symmetric signal modal field profiles after the reverse taper. All plotted field components are $|e^x|$ for TE and $|e^y|$ for TM.}
	\label{fig2}
\end{figure}

To demonstrate the proposed technique, here we provide an example based on the platform of \uppercase\expandafter{\romannumeral3}–\uppercase\expandafter{\romannumeral5} semiconductors, as shown in Fig. \ref{fig2} (a). Here a typical Al$_x$Ga$_{1-x}$As waveguide is coupled with a layer of amorphous silicon (a-Si) on top. The Al$_x$Ga$_{1-x}$As core layer has a thickness of $h_{co}=0.4$ \textmu m and an aluminum concentration of $x_{core}=0.4$. The top cladding layer has a thickness of $h_{cl}=1$ \textmu m and both claddings have an aluminum concentration of $x_{cladding}=0.8$. The SiO$_2$ and a-Si layers have thicknesses of $h_g=0.18$ \textmu m, and $h_s=0.327$ \textmu m, respectively. The waveguide is etched with an etch depth and ridge width around 1.4 and 1.6 \textmu m. For simplicity, we assume that the type-II SPDC using a TE polarized continuous-wave (CW) pump at 775 nm takes place with a quasi-phase matching (QPM) period $\Lambda$ in the Al$_x$Ga$_{1-x}$As core, in which cross-polarized photons centered around 1550 nm are generated in pairs. The periodic modulation of nonlinearity for QPM can be realized by several techniques such as orientation patterning and quantum well intermixing \cite{https://doi.org/10.1002/lpor.201000008}. Instead, modal phase matching using Bragg reflection waveguides can also be implemented (See supplementary 1 for more details). The signal photon is generated in the TE antisymmetric supermode while idler in the uncoupled TM mode, with $n_{\text{eff}}$ shown in Fig. \ref{fig2} (b). Paired photons are then separated spatially using a dichroic mirror with a splitting wavelength of 1550 nm. The output photons are entangled in polarization, with the degree of entanglement determined by the temporal-spectral indistinguishability exhibited by the BWF (See supplementary 1 for more details).

As a comparison, we also consider the conventional isolated Al$_x$Ga$_{1-x}$As waveguide without the a-Si layer. The spectra of down-converted photons (projections of $\vert\psi(\omega_S,\omega_I)\vert^2$ to $\omega_S$ and $\omega_I$ axes) are shown in the upper row of Fig. \ref{fig2} (c) for a waveguide length of 2 mm. Note that $\omega_S$ and $\omega_I$ are associated with horizontal (H) and vertical (V) polarizations, respectively, according to our assumption. The asymmetric spectra are typical for weakly birefringent Al$_x$Ga$_{1-x}$As waveguides, due to the nonzero group velocity mismatch and large group velocity dispersions \cite{PhysRevA.92.013821}. The asymmetry in spectra limits polarization entanglement, with the concurrence, an entanglement monotone, calculated to be 0.65.

On the other hand, in the proposed coupled waveguide, the modal fields of the pump and idler locate largely in the Al$_x$Ga$_{1-x}$As core and are wavelength independent. However, the location of the signal field moves from the Al$_x$Ga$_{1-x}$As core to the a-Si layer with the increase of signal wavelength, as shown in Fig. \ref{fig2} (d). As a result, the overlap factor $1/A(\omega_S,\omega_I)$ is wavelength-dependent, acting as a short-pass filter for the signal and long-pass filter for the idler, as can be seen in the middle row of Fig. \ref{fig2} (c). The sharpness of such a filter depends on the wavelength range in which the modal field is transferred from one core to the other, subject to specific waveguide designs. In this design, it is 10 nm, and the coupling wavelength is $\lambda_c=1584$ nm. The "output" spectra are shown in the lower row of Fig. \ref{fig2} (c). The concurrence in this design is increased to 0.81 due to the improved spectral overlap, and can be further increased to 0.9 by adding off-chip compensation.

Fig. \ref{fig2} (e) shows the dependence of concurrence on the coupling wavelength $\lambda_c$, which depends on the thickness of the a-Si layer. The concurrence does not improve monotonically with the decrease of $\lambda_c$, as it approaches the degenerate wavelength. This is because the filtering caused by modal overlap is single-sided for both signal and idler in this design, and thus there exists an optimal $\lambda_c$ for which the spectral overlap is maximized. For better spectral indistinguishability, one could apply this technique together with dispersion engineering that reduces the group birefringence \cite{PhysRevA.85.013838}, or design a waveguide with two pairs of coupled modes to achieve spectral filtering on both sides, which will be elaborated further in the following.

It needs to be pointed out that although we choose only to use the antisymmetric supermode in this structure, the symmetric supermode will take part in another SPDC process, generating unwanted photon pairs in the Al$_x$Ga$_{1-x}$As core with $\lambda_S>\lambda_c$, as shown in Fig. \ref{fig2} (f) (See Supplement 1 for more details). Fortunately, they can be spatially removed after an adiabatic taper section. Fig. \ref{fig2} (g) shows the modal profiles of signal in antisymmetric and symmetric supermodes after a reverse taper. Through the adiabatic process, the antisymmetric mode with $\lambda_S<\lambda_c$ stays in Al$_x$Ga$_{1-x}$As, while the symmetric mode with $\lambda_S>\lambda_c$ moves to a-Si, allowing spatial separation. In addition, the antisymmetric mode becomes more concentrated in the Al$_x$Ga$_{1-x}$As core, which benefits output coupling. Thus the coupled waveguide and the following taper form a complete source of biphoton generation and engineering. Note that the taper is used to separate biphotons that are already shaped in the generation stage, and is fundamentally different from conventional post-generation filters. 

\begin{figure}[bt!]
	\centering\includegraphics[width=5.1cm]{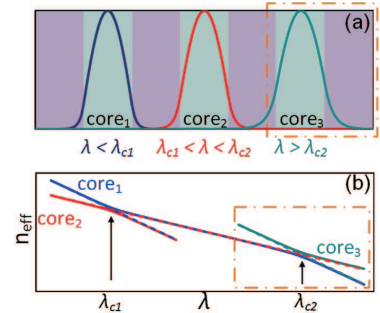}
	\caption{(a) Illustration of the composite waveguide with three core layers. Modal profiles of the coupled mode of interest at different wavelengths are superimposed on the structure. (b) The $n_{\text{eff}}$ of all three supermodes as functions of wavelength.}
	\label{fig3}
\end{figure}

As we have mentioned, our method can be extended from a single pair of coupled modes to two or more pairs of coupled modes. We again use slab waveguides for illustration. Another core layer, core$_3$, is added on the composite waveguide as shown in Fig. \ref{fig3} (a), and its corresponding modal index is $n_3$ if isolated. Layer core$_3$ may have the same material but a different width with core$_1$. As a result, $n_1$ and $n_3$ intersect with $n_2$ at slightly different wavelengths of $\lambda_{c1}$ and $\lambda_{c2}$ respectively, forming two pairs of coupled modes. The effective indices of the coupled modes are shown in Fig. \ref{fig3} (b). Note that the symmetric mode between core$_1$ and core$_2$ is also the antisymmetric mode between core$_2$ and core$_3$. Its modal field locates primarily in core$_2$ only if $\lambda_{c1}<\lambda<\lambda_{c2}$, but moves to core$_1$ if $\lambda<\lambda_{c1}$, and to core$_3$ if $\lambda>\lambda_{c2}$, as shown in Fig. \ref{fig3} (a). Therefore, if the signal is produced in this particular mode while pump and idler photons are both in core$_2$, the BWF will undergo an effective bandpass filtering on its signal. 

\begin{figure}[bt!]
	\centering\includegraphics[width=\columnwidth]{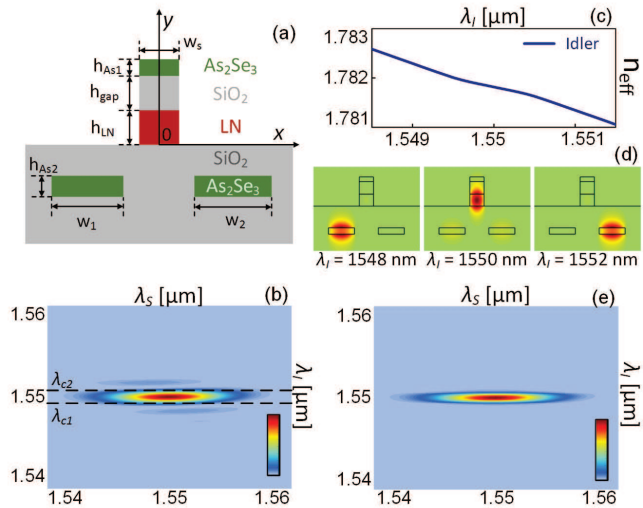}
	\caption{(a) Schematic cross-sectional view of a coupled waveguide based on thin film LN-on-insulator. (b) JSI of photon pairs for the As$_2$Se$_3$-LNOI waveguide without buried As$_2$Se$_3$ strips. The dashed lines mark the coupling wavelengths in the new design. (c) The $n_{\text{eff}}$ of idler supermode. (d) Modal profiles of the idler at three wavelengths. (e) JSI of photon pairs for the modified coupled waveguide.}
	\label{fig4}
\end{figure}

Here we provide a design example based on a different platform of LNOI that applies this technique to enhance the spectral purity of heralded single photons. The structure is based on an As$_2$Se$_3$-LNOI waveguide we developed earlier which utilizes a pair of asymmetrically coupled waveguides to achieve group velocity matching (GVM) between TE polarized signal and pump in the first place \cite{Ding:21}. Here two strips of buried As$_2$Se$_3$ with a thickness of 257 nm and slightly different widths of 3.36 \textmu m and 3.71 \textmu m are added to the original design, as shown in Fig. \ref{fig4} (a). The center positions of the As$_2$Se$_3$ strips are $x_c=\pm3.5$ \textmu m and $z_c=-1$ \textmu m. The rest design parameters can be found in \cite{Ding:21}. The waveguide is assumed to be 16 mm long and pumped by 500 fs Gaussian pulses. 

For comparison, in the original structure without buried As$_2$Se$_3$, the joint spectral intensity (JSI), defined as ${\vert\psi(\omega_S,\omega_I)\vert}^2$, looks like a horizontal ellipse, as shown in Fig. \ref{fig4} (b). The Schmidt number, a measure of spectral correlation, is 1.07. The sidelobes in JSI are the main limiting factor of the spectral purity. In the new design, the two buried As$_2$Se$_3$ cores are used to achieve coupling with the TM polarized idler mode in LN, with the $n_{\text{eff}}$ shown in Fig. \ref{fig4} (c). The modal profiles of the coupled mode at different wavelengths are shown in Fig. \ref{fig4} (d). The spatial overlap factor $1/A(\omega_S,\omega_I)$ is a function of $\omega_I$, which acts as a bandpass filter for the idler. The passband is marked by the region between the dashed lines in Fig. \ref{fig4} (b). The sidelobes are then removed from the "output" JSI as shown in Fig. \ref{fig4} (e), and the Schmidt number is further improved to 1.004. Similarly, the unwanted supermodes need to be removed by two adiabatic tapers (See supplementary 1 for more details).

\label{sec4}

It is important to note that in this example, mode coupling is applied to both polarizations but for different purposes, i. e., to achieve GVM and spectral filtering, respectively. For TE polarization, mode coupling occurs within a wavelength range significantly larger than the spectral bandwidth of the BWF. As such, the spatial overlap can be considered independent on $\omega_S$. We could, instead, implement two pairs of mode coupling for TE polarization to achieve bandpass filtering as well, enabling on-chip spectral filtering for both polarizations. Similarly, we could achieve long-pass filtering for one polarization and short-pass filtering for the other, avoiding accidental counts due to asymmetric filtering, as in the Al$_x$Ga$_{1-x}$As-a-Si example. This example shows the flexibility of asymmetric heterogeneously coupled waveguides in biphoton engineering.

Finally, we note the fabrication tolerance varies by design. For the Al$_x$Ga$_{1-x}$As-a-Si structure with a large size, with the ridge width changed by 10$\%$, the concurrence only decreases mildly to 0.76. On the other hand, for the sub-wavelength structures in the As$_2$Se$_3$-LNOI, the tolerance is much lower. For example, if the widths of two buried As$_2$Se$_3$ both increase by 1$\%$, the transmission peak of the narrow bandpass filter will shift to block most photons generated; if the difference between the two strips' widths is increased by 2$\%$, the transmission bandwidth will double and most of the sidelobes will pass through. However, by reducing the waveguide length by 5 mm, the corresponding numbers will be increased to 3$\%$ and 6$\%$, respectively, indicating a tolerance comparable to dispersion engineered nanowire waveguides of the same length \cite{Kang:14}.

In conclusion, we have developed a versatile strategy to shape photon pairs generated by SPDC using the modal field overlap as a new degree of freedom. This is achieved in asymmetric heterogeneously coupled waveguides, in which the three-field overlap could be designed such that it varies significantly with wavelengths. Combined with following tapers, intrinsic filtering of photon pairs is achieved on-chip without actual integrated filters during the generation process as opposed to filtering after generation. Design examples based on two different platforms aimed at the generation of maximal polarization entangled photons and pure heralded single photons are provided. This technique is generic and can be applied to waveguides of different material systems, providing new possibilities in photonic quantum state engineering.

\section*{Funding}
National Natural Science Foundation of China (61705053); China Postdoctoral Science Foundation (2016M600249), Heilongjiang Postdoctoral Special Funds; Fundamental Research Funds for the Central Universities, Natural Sciences and Engineering Research Council of Canada.

\section*{Disclosures}

The authors declare no conflicts of interest.

\section*{Supplemental document}
See Supplement 1 for supporting content.

\bibliography{ref}
\bibliographyfullrefs{ref}

\end{document}